\documentclass{aastex} 
\usepackage{spr-astr-addons}
\usepackage{hyperref}
\usepackage{url}
\usepackage{lscape}

\newcommand{\myemail}{rsharma@iucaa.ernet.in}

\shorttitle{Gravitational collapse of a circularly symmetric star}
\shortauthors{Sharma \textit{et al}}

\begin{document}

\title{Gravitational collapse of a circularly symmetric star in an anti-de Sitter spacetime}

\author{Ranjan Sharma\altaffilmark{1,a}, Shyam Das\altaffilmark{1,b}, Farook Rahaman\altaffilmark{2,c} and Gopal Chandra Shit\altaffilmark{2,d}} 

\altaffiltext{1}{Department of Physics, P. D. Women's College, Club Road, Jalpaiguri 735101, India.}
\altaffiltext{a}{\myemail}
\altaffiltext{b}{dasshyam321@gmail.com}

\altaffiltext{2}{Department of Mathematics, Jadavpur University, Kolkata 700032, West Bengal, India.}
\altaffiltext{c}{rahaman@iucaa.ernet.in}
\altaffiltext{d}{gopal$\_$iitkgp@yahoo.co.in}

\begin{abstract}
We investigate the collapse of a circularly symmetric star with outgoing radiation in ($2+1$)-dimensional anti-de Sitter spacetime. The exterior spacetime of the collapsing star is assumed to be described by the non-static generalization of the Ba\~{n}ados, Teitelboim and Zanelli [{\em Phys. Rev. Lett.} {\bf 69} (1992) 1849 ] metric. Making use of the junction conditions joining smoothly the interior and the exterior spacetimes across the boundary, we analyze the impacts of various factors on the evolution of the star which begins its collapse from an initial static configuration. In particular, depending on initial conditions, two possible outcomes of the collapse process are shown: (i) formation of a BTZ black hole; and (ii) evaporation of all mass-energy even before the singularity is reached.
\end{abstract}

\keywords{($2+1$)-dimensional gravity; gravitational collapse; exact solution; BTZ spacetime.}

\section{\label{sec1}Introduction}  

One of the most fundamental problems in general relativity is the successful prediction of the final stage of a gravitationally collapsing system. During the collapse of a massive star matter can reach extreme conditions in which nothing can stop the collapse. The general relativistic prediction in such a situation is the formation of a spacetime singularity. The state of singularity may be a naked one in the sense that light rays emanating from it can reach any far-away observer or a black hole implying that light rays from it may be trapped within its event horizon \citep{Joshibook}. As of now, our understanding about the final outcome of a gravitationally collapsing object is known to rely on a couple of adopted conjectures only \citep{Joshibook,Thornebook}. Understanding the characteristics and subsequent end stage of a collapsing system  is not just important from the theoretical point of view, it has tremendous observational consequences as well \citep{Vir1,Vir2}. Though quantum theories are expected to play a dominant in the late stages of a collapsing matter, the theoretical framework of quantum gravity remains unaccomplished till date. Consequently, in the absence of any physical or mathematical theory governing gravitational collapse, studies of collapsing systems composed of a wide variety of matter distributions with or without dissipation in different background spacetimes within the realm of classical gravity are expected to play a key role in promoting our understanding about the nature and final outcome of the collapsing system. 

Realistic modelling of a collapsing matter source in classical gravity is, however, extremely difficult due to highly non-linear nature of the governing field equations. Various simplifying techniques are often adopted to generate models of collapsing systems. Studies of a highly idealized dust cloud collapsing under the influence of its own gravity was a first step in this direction \citep{OppenS}. Later, \cite{Vaidya} metric describing the exterior gravitational field of a radiating fluid sphere and subsequent junction conditions obtained by \cite{Santos} have enabled many investigators to construct and analyze a large variety of collapsing systems (see e.g., \cite{Joshi} and references therein). 

Mathematical complexities involved in a generally covariant four dimensional gravitating system have prompted many investigators to carry out similar of studies in lower dimensions as well. In fact, ($2+1$)-dimensional gravity has became extremely fascinating with the discovery of the black hole solution obtained by Ba$\tilde{n}$ados, Teitelboim and Zanelli (henceforth BTZ) \citep{BTZ}. The BTZ metric is an exact solution of the Einstein's field equations describing the exterior gravitational field of a circularly symmetric Einstein-Maxwell system in the presence of a negative cosmological constant and is characterized by its mass, angular momentum and charge \citep{Carlip}. 

The fact that (in the presence of a negative cosmological constant) there exists a black hole solution in ($2+1$)-dimensional gravity has inspired many investigators to develop and study equilibrium configurations of circularly symmetric self-gravitating systems in anti-de Sitter spacetimes. In particular, interior solutions corresponding to exterior BTZ spacetime have been developed and analyzed by \cite{Cruz}, \cite{Cruz2}, \cite{Berredo}, \cite{Paulo}, \cite{Sharma}, \cite{Garc}, \cite{Ayan,Ayan2}, \cite{Farook}, \cite{Cataldo} and \cite{Garcia}. Recently, a non-commutative geometry inspired $3$-dimensional charged black hole solution has been reported \citep{Farook2015}.  

The objective of the current investigation is to construct a model of a circularly symmetric collapsing matter source in an anti-de Sitter background spacetime and analyze its subsequent evolution. Earlier, \cite{RossMann} have investigated the circumstances under which a gravitationally collapsing dust cloud would form a black hole in ($2+1$)-dimensional spacetime. By considering the three-dimensional analogue of Oppenheimer-Snyder collapse, they have shown that a stationary BTZ black hole might be a natural consequence of the collapsing dust in the presence of a negative cosmological constant. Later, \cite{MannJohn} have demonstrated that a collapsing dust shell might not necessarily end to a black hole. By incorporating a cosmological constant term, \cite{Gutti} have analyzed ($2+1$)-dimensional  gravitational collapse of a spherically symmetric inhomogeneous dust and discussed the nature of singularity which was found to depend on the values of the cosmological constant. \cite{Gutti2} have investigated the creation of quantum particles during the collapse of a circularly symmetric dust with or without a cosmological constant term. Self-similar solutions corresponding to the collapse of a circularly symmetric anisotropic fluid have been proposed by \cite{Martins}. Gravitational collapse of a self-similar perfect fluid in ($2+1$)-dimensions has also been analyzed by \cite{Miguelote} where it has been shown that both black holes and naked singularities might be the possible end states of collapse. \cite{Chan2} have investigated the gravitational collapse of a massless scalar field in the presence of a negative cosmological constant in $3$-dimensions. The critical behaviour of a massless scalar field near the black hole threshold has been analyzed by \cite{Choptuik1} and a mass scaling law for the formation of the black hole has been proposed. To test the CCC, \cite{Ghosh} has developed a new class of exact non-static charged BTZ-like solutions in ($n+1$)-dimensions describing the gravitational collapse of a charged null fluid in an anti-de Sitter spacetime. 

In this work, we plan to develop model of a circularly symmetric collapsing matter source by specifying the appropriate interior and exterior spacetimes and solving the relevant junction conditions. Note that in ($3+1$) dimensions, \cite{Vaidya} metric appropriately describes the exterior gravitational field of a collapsing matter cloud with outgoing radiation. In ($3+1$)-dimensional anti-de Sitter spacetime, \cite{Lemos} has studied the process of gravitational collapse with outflowing radiation. By introducing a cosmological constant term in the \cite{Vaidya} metric, \cite{Govender} have investigated the nature of collapse of a spherically symmetric matter source with dissipation. In the braneworld scenario, while investigating the collapse of a the null fluid which eventually collapses onto a flat Minkowski cavity, \cite{Dadhich} have obtained a solution which is analogous to the Vaidya solution. The ($2+1$)-dimensional analogue of Vaidya metric describing the exterior gravitational field of a collapsing null fluid in the presence of a negative cosmological constant, which is also a non-static generalization of the uncharged and spin-less BTZ metric, has been proposed independently by \cite{Virbhadra} and \cite{Hussain1,Hussain2}. For the collapse of a radiating matter source admitting an EOS of the form $p= k\rho~(k\leq 1)$, \cite{Hussain2} has shown that the BTZ solution might be regained as a special case for $k=1$ while for $k < 1$, it would generate black hole solutions with multiple apparent horizons. Assuming the background to be the non-static generalization of the BTZ solution, \cite{Vagenas} has analyzed the energy and momentum distributions associated with a non-static circularly symmetric spacetime. In our construction, we assume that the exterior region of the collapsing matter is described by the ($2+1$)-dimensional analogue of Vaidya metric proposed and analyzed by \cite{Virbhadra} and \cite{Hussain1,Hussain2}. 

The paper has been organized as follows. In Sec.~\ref{sec2}, the field equations for a circularly symmetric collapsing star in an anti-de Sitter spacetime have been laid down. The junction conditions joining smoothly the interior and exterior spacetimes have been obtained in Sec.~\ref{sec3}. The junction conditions yield a non-linear differential equation which governs the overall evolution of the star. We have solved the temporal equation in Sec.~\ref{sec4}. Making use of a solution corresponding to an initial static configuration, we have generated a specific collapsing model in Sec.~\ref{sec5}. Physical processes during the evolutionary stages of the collapsing star have been analyzed in Sec.~\ref{sec6} and some concluding remarks have been made in Sec.~\ref{sec7}.

\section{\label{sec2} Field equations}

In ($3+1$)-dimensions, the \cite{Vaidya} solution describing the exterior spacetime of a spherically symmetric collapsing matter with outgoing radiation in the presence of a cosmological constant has the form
\begin{eqnarray}
ds^2_{+} &=& -\left(1-\frac{2m(u)}{\tilde{r}}-\frac{\Lambda \tilde{r}^2}{3}\right)du^2-2 du d\tilde{r} \nonumber\\
&&+\tilde{r}^2(d\theta^2+\sin\theta^2d\phi^2).\label{vm} 
\end{eqnarray}
The ($2+1$)-dimensional analogue of the metric (\ref{vm}) representing the exterior of a non-static circularly symmetric source in the presence of a negative cosmological constant and null fluid, reported independently by \cite{Virbhadra} and \cite{Hussain1,Hussain2}, has the form   
\begin{equation}
ds^2_{+} = -(-m(u)-\Lambda \tilde{r}^2)du^2-2 du d\tilde{r}+\tilde{r}^2 d\theta^2,\label{virm} 
\end{equation}
in system of coordinates ($u,\tilde{r},\theta$). By introducing a coordinate transformation
$$u = t-\int {\frac{d\tilde{r}}{-m(u)-\Lambda \tilde{r}^2}}$$ and setting $m(u)= M_0$, the metric (\ref{virm}) can be reduced to the well-known spin-zero BTZ metric 
\begin{equation}
ds^2_{BTZ} = -(-M_0-\Lambda \tilde{r}^2)dt^2+\frac{d\tilde{r}^2}{-M_0 -\Lambda \tilde{r}^2} + \tilde{r}^2 d\theta^2,\label{btzm} 
\end{equation}
describing a BTZ star whose event horizon is located at $\tilde{r} = \sqrt{-\frac{M_0}{\Lambda}}$. In our construction, we assume that the collapse begins from an initial configuration of radius $r_0  >  \sqrt{-\frac{M_0}{\Lambda}}$ and the exterior region of the subsequent collapsing star is described by the metric (\ref{virm}). We couch the interior spacetime of the collapsing star in the form
\begin{equation}
ds_{-}^2 = - e^{2\nu(r,t)}dt^2 + e^{2\mu(r,t)}dr^2 + r^2 s^2(t)d\theta^2,\label{intm}
\end{equation}
where $\nu(r,t)$, $\mu(r,t)$ and $s(t)$ are yet to be specified. The energy-momentum tensor of the material composition filling the interior of the collapsing star with outgoing radiation is assumed to be 
\begin{equation}
T_{\alpha\beta} = (\rho + p_t)u_{\alpha} {u_\beta} + p_t g_{\alpha \beta} + (p_r - p_t)\chi_\alpha \chi_\beta + \epsilon l_\alpha l_\beta,\label{emt}
\end{equation}
where, $\rho$ represents the energy density, $p_t$ and $p_r$ represent tangential and radial fluid pressures, respectively and $\epsilon$ is the energy density of the outflowing radiation. In (\ref{emt}), $\chi^\alpha$ is the unit $3$-vector along the radial direction, $l^\alpha$ is a radially directed null $3$-vector and $u^\alpha$ is the $3$-velocity of the fluid satisfying the following relations:
\begin{center}
$u^\alpha u_\alpha = -1,$~~~$l^\alpha u_\alpha = -1,$~~~$l^\alpha l_\alpha = 0,$~~~$\chi^\alpha \chi_\alpha = 1,$\\
$\chi^\alpha u_\alpha = 0,$~~~$u^\alpha = e^{-\nu} \delta_t^\alpha,$~~~$l^\alpha = e^{-\nu} \delta_t^\alpha+ e^{-\mu} \delta_r^\alpha.$
\end{center}

The Einstein's field equations (we set $G=1$ and $c = 1$)
\begin{equation} 
R_{\alpha\beta} -\frac{1}{2} g_{\alpha\beta} R = 2\pi T_{\alpha\beta} - \Lambda g_{\alpha\beta},\label{efq}
\end{equation}
then yield a system of the following four independent equations: 
\begin{eqnarray}
2\pi  (\rho+ \epsilon) +\Lambda &=& \frac{\dot{s}\dot{\mu}e^{-2\nu}}{s}+\frac{\mu'e^{-2\mu}}{r},\label{efq1}\\
2\pi (p_r+ \epsilon)-\Lambda &=& \frac{\nu' e^{-2\mu}}{r}+\frac{e^{-2\nu}\{-\ddot{s}+\dot{s}\dot{\nu}\}}{s},\label{efq2}\\
2\pi  p_t  -\Lambda &=& e^{-2\nu}\{{\dot{\mu}(\dot{\nu}-\dot{\mu})-\ddot{\mu}}\}\nonumber\\
&& +e^{-2\mu}\{{-\nu'\mu'+\nu'^2+\nu''}\},\label{efq3}\\
2 \pi  \epsilon &=& -\frac{s\dot{\mu}+\dot{s}\{-1+r\nu'\}}{r s}e^{-(\nu+\mu)}.\label{efq4} 
\end{eqnarray}
Physical behaviour of the collapsing star can be understood by solving the above set of equations.

\section{\label{sec3}Junction conditions}

In order to obtain the necessary junction conditions joining the interior $V^{-}$ and exterior $V^{+}$ spacetimes, we follow the prescription of \cite{Israel}, which demands that the first and second fundamental forms of the interior and exterior spacetimes be continuous across the junction hyper-surface $\Sigma$, a time-like matching surface dividing the two distinct regions of the collapsing system. If $g_{ij}$ be the intrinsic metric to $\Sigma$ so that 
\begin{equation}
ds^2_{\Sigma} = g_{ij}d\xi^i d\xi^j,~~~~~i=1,2\label{intrinm1}
\end{equation} 
and $g^\pm_{\alpha\beta}$ be the metric corresponding to $V^\pm$ so that
\begin{equation}
ds^2_\pm = g^\pm_{\alpha\beta}d\chi^\alpha_\pm d\chi^\beta_\pm,~~~~~\alpha=0,1,2\label{intrinm2}
\end{equation}
then the first junction condition is obtained from the following requirement:
\begin{equation}
(ds^2_{-})_\Sigma = (ds^2_{+})_\Sigma = ds^2_\Sigma.\label{eqj1} 
\end{equation}
The second junction can be obtained by matching the extrinsic curvatures:
\begin{equation}
(K^+_{ij})_{\Sigma} = (K^-_{ij})_{\Sigma},\label{eqj2}
\end{equation}
where,
\begin{equation}
K^\pm_{ij} = -n^\pm_\alpha\frac{\partial^2\chi^\alpha_\pm}{\partial \xi^i\partial \xi^j}-n^\pm_\alpha\Gamma^\alpha_{\mu\nu}\frac{\partial\chi^\mu_\pm}{\partial\xi^i}\frac{\partial\chi^\nu_\pm}{\partial\xi^j},\label{eqj3}
\end{equation}
In (\ref{eqj3}), $n^\pm_\alpha$ are the components of the normal vector to $\Sigma$ in $\chi^\alpha_{\pm}$ coordinates. We express the intrinsic metric to $\Sigma$ as
\begin{equation}
ds^2_{\Sigma} = g_{ij}d\xi^i d\xi^j = -d\tau^2+\Re^2(\tau)d\theta^2.\label{eqj4} 
\end{equation}
The junction condition (\ref{eqj1}) for the intrinsic metric (\ref{eqj4}) corresponding to the interior spacetime (\ref{intm}) yields 
\begin{eqnarray}
e^{\nu(r_\Sigma,t)}\frac{dt}{d\tau} &=& 1,\label{eqj5}\\
r_\Sigma s(t) &=& \Re(\tau).\label{eqj6} 
\end{eqnarray}
For the interior region we have
\begin{equation} 
f(r,t) = r-r_\Sigma = 0.\label{eqj7} 
\end{equation}
The vector with components $\frac{\partial f}{\partial \chi^\alpha_-}$ is orthogonal to $\Sigma$ and, therefore, the unit normal to $\Sigma$ in coordinates $\chi^\alpha_-$ is obtained as
\begin{equation} 
n^-_\alpha = \{0,e^{\mu(r_\Sigma,t)},0\}.\label{eqj8}
\end{equation}
The extrinsic curvatures corresponding to the interior region are then obtained as
\begin{eqnarray}
K^-_{\tau\tau} &=& -e^{-\mu(r,t)}\nu'(r,t),\label{eqj9}\\
K^-_{\theta\theta} &=& e^{-\mu(r,t)}r s^2(t),\label{eqj10}
\end{eqnarray}
The junction condition (\ref{eqj1}) for the intrinsic metric (\ref{eqj4}) corresponding to the exterior space-time (\ref{virm}) yields
\begin{eqnarray}
\tilde{r}_\Sigma(u) &=& \Re(\tau),\label{eqj11}\\
\left(\frac{du}{d\tau}\right)^{-2}_\Sigma &=& \left(2 \frac{d\tilde{r}_\Sigma}{du}-m(u)-\Lambda \tilde{r}_\Sigma^2\right).\label{eqj12}
\end{eqnarray}
Combining (\ref{eqj6}) and (\ref{eqj11}), we get
\begin{equation}
r_\Sigma s(t) = \tilde{r}_\Sigma(u),\label{eqj13}
\end{equation}

For the exterior region, we have 
\begin{equation}
f(\tilde{r},u) = \tilde{r}-\tilde{r}_\Sigma(u) = 0,\label{eqj14} 
\end{equation}
and
\begin{equation}
\frac{\partial f}{\partial\chi^\alpha_+} = \left(-\frac{d\tilde{r}_\Sigma}{du},1,0\right).\label{eqj15}
\end{equation} 
The unit normal to $\Sigma$ takes the form
\begin{equation}
n^+_\alpha = \left(2 \frac{d\tilde{r}_\Sigma}{du}-m(u)-\Lambda \tilde{r}_\Sigma^2\right)^{-\frac{1}{2}}\left(-\frac{d\tilde{r}_\Sigma}{du},1,0\right).\label{eqj16}
\end{equation}
Combining Eqs.~(\ref{eqj12}) and (\ref{eqj16}), we get  
\begin{equation}
n^+_\alpha = \left(-\frac{d\tilde{r}_\Sigma}{d\tau},\frac{du}{d\tau},0\right).\label{eqj17}
\end{equation}
The extrinsic curvatures corresponding to the exterior region are then obtained as
\begin{eqnarray} 
K^+_{\tau\tau} &=& \left(\frac{du}{d\tau}\right)\left[-\left(\frac{d^2\tilde{r}}{d \tau^2}\right)+3\Lambda\tilde{r}\left(\frac{d\tilde{r}}{d\tau}\right)\left(\frac{du}{d\tau}\right)\right. \nonumber\\
&&\left. -\frac{1}{2}\left(\frac{du}{d\tau}\right)^2\left\{2\Lambda^2\tilde{r}^3+2\Lambda \tilde{r} m(u)-\frac{dm(u)}{du}\right\}\right.\nonumber\\
&&\left.+\left(\frac{d\tilde{r}}{du}\right)\left(\frac{d^2u}{d\tau^2}\right)\right]_\Sigma,\label{eqj18}\\
K^+_{\theta\theta} &=& \left[\tilde{r}\left(\frac{d\tilde{r}}{d\tau}\right)-\left(\frac{du}{d\tau}\right)(m(u)+\Lambda\tilde{r}^2)\tilde{r}\right]_\Sigma.\label{eqj19}
\end{eqnarray}
Continuity of the extrinsic curvature parameters
\begin{equation}
(K^+_{\theta\theta})_\Sigma  = (K^-_{\theta\theta})_\Sigma,\label{eqj20}
\end{equation} 
yields 
\begin{eqnarray} 
e^{-\mu(r_\Sigma,t)}r_\Sigma s^2(t) &=& \left[\tilde{r}\left(\frac{d\tilde{r}}{d\tau}\right)-\right.\nonumber\\
&&\left.\left(\frac{du}{d\tau}\right)\left(m(u)+\Lambda\tilde{r}^2\right)\tilde{r}\right]_\Sigma.\label{eqj21}
\end{eqnarray}
Using Eqs.~(\ref{eqj6}) and (\ref{eqj11}), we rewrite Eq.~(\ref{eqj21}) as
\begin{equation}
e^{-\mu(r_\Sigma,t)}s(t) = \left[\left(\frac{d\tilde{r}}{d\tau}\right)-\left(\frac{du}{d\tau}\right)(m(u)+\Lambda\tilde{r}^2)\right]_\Sigma.\label{eqj22}
\end{equation}
Continuity of the extrinsic curvature parameters
\begin{equation}
(K^+_{\tau\tau})_\Sigma = (K^-_{\tau\tau})_\Sigma,\label{eqj23} 
\end{equation}
yields
\begin{eqnarray}
-e^{-\mu(r_\Sigma,t)}\nu'(r_\Sigma,t) = \left(\frac{du}{d\tau}\right)\left[-\left(\frac{d^2\tilde{r}}{d \tau^2}\right)+3\Lambda\tilde{r}\left(\frac{d\tilde{r}}{d\tau}\right)\left(\frac{du}{d\tau}\right)\right.\nonumber\\
\left.-\frac{1}{2}\left(\frac{du}{d\tau}\right)^2\left\{2\Lambda^2\tilde{r}^3+2\Lambda \tilde{r} m(u)-\frac{dm(u)}{du}\right\}\right.\nonumber\\
\left.+\left(\frac{d\tilde{r}}{du}\right)\left(\frac{d^2u}{d\tau^2}\right)\right]_\Sigma.\label{eqj24}
\end{eqnarray}
Using Eqs.~(\ref{eqj5}), (\ref{eqj6}), (\ref{eqj11}), (\ref{eqj12}), (\ref{eqj13}) and (\ref{eqj22}), the mass function is then obtained as
\begin{eqnarray} 
m(u) \stackrel{\Sigma}{=}  m(r_{\Sigma}, t) &=& \left[\dot {s}^2 r_\Sigma^2 e^{-2\nu(r_\Sigma,t)}\right.\nonumber\\
&&\left. - s^2 e^{-2\mu(r_\Sigma,t)}-\Lambda r_\Sigma^2 s^2\right],\label{eqj25}
\end{eqnarray}
where, $ m(r_{\Sigma}, t)$ is the total mass within a boundary $r=r_{\Sigma}$ at any instant $t$.

From (\ref{eqj24}), we obtain
\begin{eqnarray}
\left[-e^{-\mu}{\nu}'\right]_{\Sigma} =\nonumber\\
 \left[\frac{e^{-\nu}\left[e^{\nu}(-\dot{s} + s\dot{\mu})+e^{\mu}r(e^{2\nu}\Lambda s-\ddot{s}+\dot{s}\dot{\nu})\right]}{e^{\nu} s + e^{\mu}r\dot{s}}\right]_{\Sigma},\label{eqj26}
\end{eqnarray}
which on rearrangement takes the form
\begin{eqnarray}
\Lambda + \frac{\nu' e^{-2\mu}}{r}+\frac{e^{-2\nu}\{-\ddot{s}+\dot{s}\dot{\nu}\}}{s}+\nonumber\\
e^{-(\mu+\nu)}\left\{\frac{s\dot{\mu}+\dot{s}\{-1+r\nu'\} }{r s}\right\} \stackrel{\Sigma}{=} 0.\label{eqj27}
\end{eqnarray}
Using Eqs.~(\ref{efq2}) and (\ref{efq4}) in (\ref{eqj27}), we obtain
\begin{equation}
p_r \stackrel{\Sigma}{=} 0.\label{eqj28}
\end{equation}
To generate a dynamical model, we need to determine $s(t)$ which can be obtained by solving the junction condition (\ref{eqj28}).

\section{\label{sec4} Determination of $s(t)$}

We assume that the metric potentials in Eq.~(\ref{intm}) are separable in their variables and accordingly we write
\begin{equation}
e^{\nu(r,t)} = e^{\nu_0(r)},~~~~~e^{\mu(r,t)} = e^{\mu_0(r)}s(t),\label{eqstm}
\end{equation}
where the time-dependent part of the metric potential $g_{tt}$ gets absorbed. The advantage of the above assumption is that one can regain the interior spacetime of a static circularly symmetric BTZ star described by the metric potentials ($\nu_0(r)$, $\mu_0(r)$) simply by setting $s = 1$. Combining Eqs.~(\ref{efq2}) and (\ref{efq4}) and making use of the junction condition (\ref{eqj28}), we then have 
\begin{equation} 
s\ddot{s} - a\dot{s} - b s^2 - c = 0.\label{sureq}
\end{equation}
where, 
$$a = \left[\nu'_0 e^{(\nu_0-\mu_0)}\right]_{\Sigma},\\
b = \left[\Lambda e^{2\nu_0}\right]_{\Sigma},\\
c = \left[\nu'_0 e^{2(\nu_0-\mu_0)}/r\right]_{\Sigma}$$
are constants evaluated at the surface $\Sigma$. Eq.~(\ref{sureq}) governs the temporal behaviour of the collapsing matter.

Solution of Eq.~(\ref{sureq}) in closed form is not available. If we set $e^{\gamma_0(r)} = 1$, Eq.~(\ref{sureq}) reduces to 
\begin{equation}
\ddot{s} -b s = 0,\label{sureq1}
\end{equation}
which admits a solution
\begin{equation}
s(t) = A_1 \cos{\sqrt{-b}t} + A_2\sin{\sqrt{-b}t},\label{seq1sol}
\end{equation}
where $A_1$ and $A_2$ are integration constants. This class of solutions can utilized to model a collapsing dust surrounded by vacuum, i.e., no radiation field as shown by \cite{RossMann}.  

For $e^{\gamma_0(r)} \neq 1$, making use of the homotopy perturbative approach, we propose an approximate solution of Eq.~(\ref{sureq}) in the form 
\begin{equation}
s(t) = c_1 +c_2 t + c_3 t^2 + .....,\label{seq2sol}
\end{equation}
where
\begin{eqnarray}
c_2 &=& \frac{(a^2+b-c) \pm \sqrt{(c-b-a^2)^2+4a^2(b+c)}}{2a},\nonumber\\
c_3 &=& \frac{1}{4}[(a^2+b-c) \pm \sqrt{(c-b-a^2)^2+4a^2(b+c)}]\nonumber\\
&&+\frac{(b+c)}{2},\nonumber
\end{eqnarray} 
and $c_1$ are constants. 

Let us assume that the collapse begins from an initial static configuration at time $t_{i}$. Then, we must have (i) $s(t_{i}) = 1$ and (ii) $\dot{s} (t_{i}) = 0$. The collapse will continue till the event horizon is formed at time $t_{h}$. 

Condition (i) yields
\begin{equation}
t_{i} =\frac{-c_2 \pm \sqrt{c_2^2-4c_3(c_1-1)}}{2c_3}.\label{eqcond1}
\end{equation}
Condition (ii) yields
\begin{equation}
t_{i} = -\frac{c_2}{2c_3}.\label{eqcond2}
\end{equation}
Combining Eqs.~(\ref{eqcond1}) and (\ref{eqcond2}), we get
\begin{equation}
c_1 = 1+\frac{c_2^2}{4c_3}.\label{eqcond3}
\end{equation}
The constants $a$, $b$ and $c$ can be conveniently fixed so as to satisfy the above requirement. In our construction, without any loss of generality, we set $c_1 = 1$ and $c_2 = 0$ satisfying the above requirements so that the solution (\ref{seq2sol}) takes the form
\begin{equation}
s(t) = 1+\frac{(b+c)}{2}t^2.\label{stsol}
\end{equation}
Eq.~(\ref{stsol}) implies that the collapse begins at $t_i =0$. For a contracting ring, we must have $\dot{s} < 0$ which can be achieved by fulfilling the condition $ b+c < 0$.

\section{\label{sec5} A particular model}

After finding the solution for $s(t)$, the problem of generating the collapsing model is now reduced to specifying a suitable solution for the metric potentials $\nu_0(r)$ and $\mu_0(r)$ describing the initial static configuration. For the initial static configuration, we take the class of solutions for a stationary circularly symmetric star in anti-de Sitter space obtained by \cite{Ayan2}. The solution has been obtained for an anisotropic  matter distribution satisfying a linear equation of state of the form $p_r = \alpha \rho +\beta$ and can be expressed as
\begin{eqnarray}
\mu_0 &=& \frac{A r^2}{2},\label{efq11}\\
\nu_0 &=& \frac{\alpha A r^2}{2} -\left(\frac{\alpha\Lambda +\Lambda -2\pi\beta}{2A}\right)e^{Ar^2} +\delta.\label{efq12}
\end{eqnarray}
In (\ref{efq11}) and (\ref{efq12}), $A$, $\alpha$, $\beta$ and $\delta$  are constants which can be determined from the appropriate boundary conditions. Matching the interior solution to the static exterior BTZ metric at the boundary of the star $r_0 > \sqrt{-\frac{M_0}{\Lambda}}$ and imposing the condition $p_r (r=r_0) = 0$, we have
\begin{eqnarray}
A &=& -\frac{1}{r_0^2}\ln(-M_0-\Lambda r_0^2),\label{sconeq1}\\
r_0 &=& \frac{1}{\sqrt{A}}\left[\ln\left(\frac{A\alpha}{\alpha \Lambda -2\pi\beta}\right)\right]^{1/2},\label{sconeq2}\\
\delta & =& \frac{\alpha\Lambda+\Lambda-2\pi\beta}{2A(-M_0-\Lambda r_0^2)}-\frac{Ar_0^2(1+\alpha)}{2}. \label{sconeq3}
\end{eqnarray}
It has been shown that the solution is well behaved and satisfies all the regularity conditions required for a physically viable model \citep{Ayan2}. As the collapse begins from the initial static configuration, the dynamical variables take the form
\begin{eqnarray}
2\pi (\rho+ \epsilon) + \Lambda &=& \frac{\dot{s}^2 e^{-2\nu_{0}}}{s^2}+\frac{\mu_0'e^{-2\mu_0}}{r s^2},\label{efq5}\\
2\pi (p_{r}+ \epsilon) - \Lambda &=& \frac{\nu_0'e^{-2\mu_0}}{r s^2}-\frac{\ddot{s}e^{-2\nu_{0}}}{s},\label{efq6}\\
2\pi  p_{t} - \Lambda &=& \frac{e^{-2\mu_{0}}(-\nu'_{0}\mu'_{0}+{\nu'_{0}}^2+\nu''_{0})}{s^2} \nonumber\\
&&-\frac{\ddot{s} e^{-2\nu_{0}}}{s},\label{efq7}\\
2\pi \epsilon &=& -\frac{\dot{s}}{s}\nu'_0 e^{-(\nu_0+\mu_0)}.\label{efq8} 
\end{eqnarray}
From Eq.~(\ref{efq8}), we note that at the onset of collapse $\epsilon = 0$ (since $\dot{s}(t_i=0) = 0$). During the evolution, $\epsilon$ will be positive if $\nu'_0$ remains positive since $\dot{s} < 0$.

Using Eq.~(\ref{eqj25}), we express the total mass of the collapsing star at a given instant $t$ within the shrinking boundary $r_{\Sigma}=r_0 s(t)$ as
\begin{equation}
m(r,t) \stackrel{\Sigma}{=} \left[\dot{s}^2 r^2 e^{-2\nu_0(r)}- e^{-2\mu_0(r)}-\Lambda r^2 s^2\right].\label{eqf19}
\end{equation}
Substituting $s=1$ and $\dot{s}=0$ in (\ref{eqf19}), we get
\begin{equation}
M_0 =  \left[- e^{-2\mu_0(r)}-\Lambda r^2\right]_{\Sigma},\label{eqf20}
\end{equation}
as in \cite{Ayan2} paper.

\section{\label{sec6} Physical analysis}

Making use of the developed model, we adopt numerical techniques to investigate the role of various factors governing the collapse process. 

\begin{figure}[tb]
\includegraphics[width=\columnwidth]{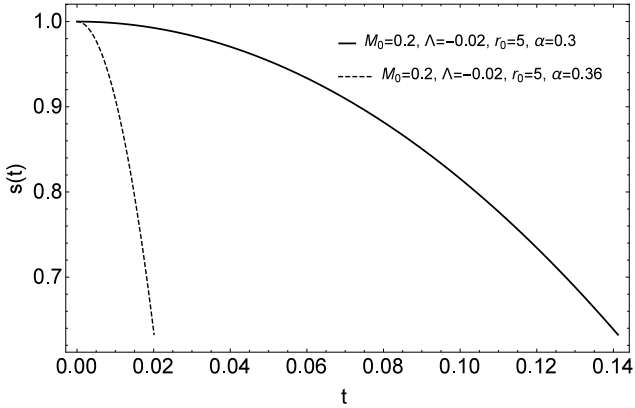}\caption{Evolution of the scale factor $s(t)$ for different values of the EOS parameter $\alpha$} \label{fg1btz}
\end{figure}

\begin{figure}[tb]
\includegraphics[width=\columnwidth]{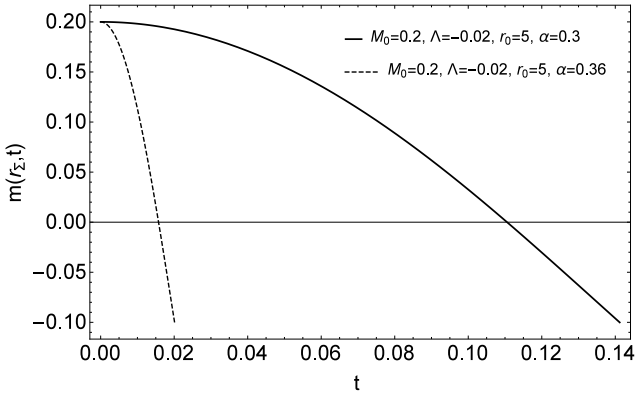}\caption{Evolution of the total mass $m(r_{\Sigma},t)$ for different values of the EOS parameter $\alpha$} \label{fg2btz}
\end{figure}

\begin{figure}[tb]
\includegraphics[width=\columnwidth]{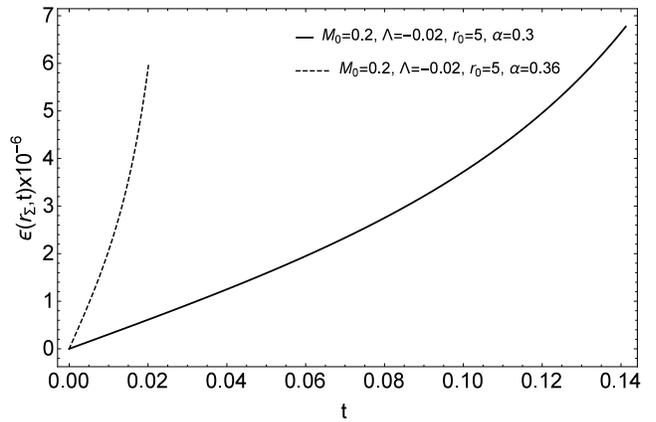}\caption{Evolution of the radiation density $\epsilon(r_{\Sigma},t)$ for different values of the EOS parameter $\alpha$}\label{fg3btz}
\end{figure}

\begin{figure}[tb]
\includegraphics[width=\columnwidth]{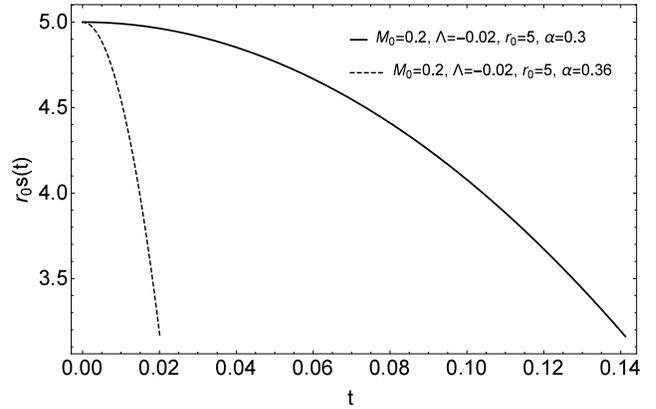}\caption{Evolution of radius $r_0 s(t)$ for different values of the EOS parameter $\alpha$}\label{fg4btz}
\end{figure}

\begin{figure}[tb]
\includegraphics[width=\columnwidth]{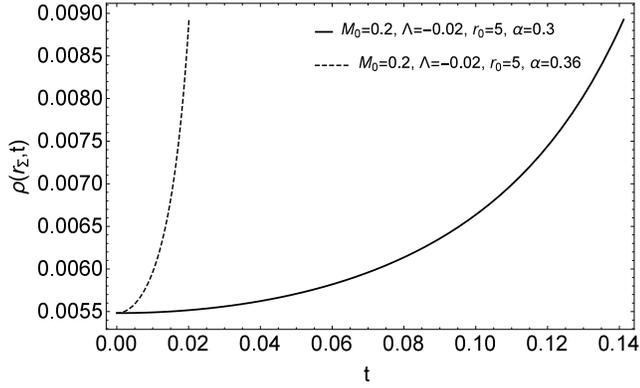}\caption{Evolution of surface density $\rho(r_{\Sigma},t)$ for different values of the EOS parameter $\alpha$} \label{fg5btz}
\end{figure}

\begin{figure}[tb]
\includegraphics[width=\columnwidth]{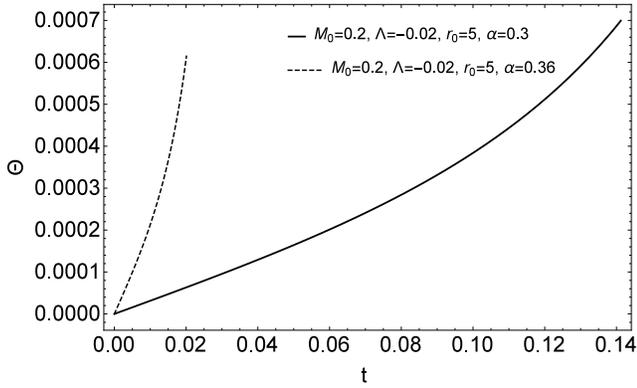}\caption{Evolution of collapse rate $\Theta$ for different values of the EOS parameter $\alpha$} \label{fg6btz}
\end{figure}

\begin{itemize}
\item {\bf Case I:}\\
Let the initial stage of the collapsing configuration is characterized by $M_0 = 0.8$ and $r_0 = 10$ and $\Lambda = -0.01$. Since three out of the four remaining model parameters ($A$, $\delta$, $\alpha$ and $\beta$) can be fixed by solving Eqs.~(\ref{sconeq1})-(\ref{sconeq3}), we are free to choose one more parameter and, accordingly, we assume $\alpha =0.1$. The constants are then obtained as $A = 0.0161$, $\beta = -0.0002$ and $\delta = 14.6982$. Using the set of values, we have shown graphically behaviour of the physical quantities. We have shown how the scale factor $s(t)$, total mass $m(r_{\Sigma}, t)$ and radius $r_0 s(t)$ decrease from their initial values in figures (\ref{fg9btz}), (\ref{fg10btz}) and (\ref{fg12btz}), respectively. The radiation density is zero initially and increases as time progresses as shown in Fig.~(\ref{fg11btz}). The energy-density evaluated at the boundary also increases with time as shown in Fig.~(\ref{fg13btz}). The horizon radius $r_h = \sqrt{-\frac{M_0}{\Lambda}} = 8.9443$, in this case, is reached at time $t_h=0.0016$ where the mass is finite and
positive implying the formation of a BTZ black hole.

\begin{figure}[tb]
\includegraphics[width=\columnwidth]{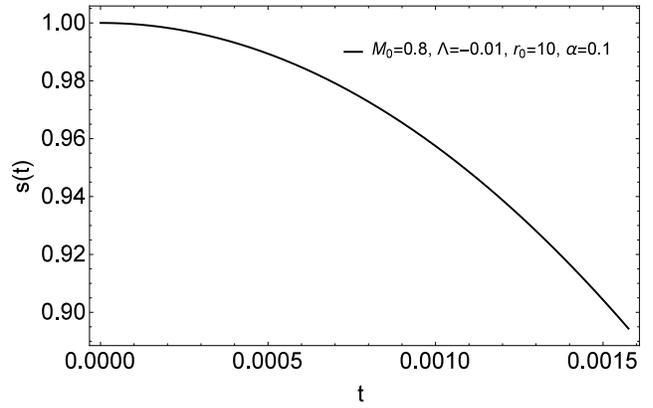}\caption{Evolution of the scale factor $s(t)$} \label{fg9btz}
\end{figure}

\begin{figure}[tb]
\includegraphics[width=0.75\columnwidth]{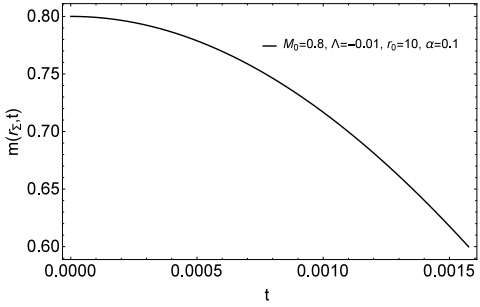}\caption{Evolution of the total mass $m(r_{\Sigma},t)$} \label{fg10btz}
\end{figure}

\begin{figure}[tb]
\includegraphics[width=\columnwidth]{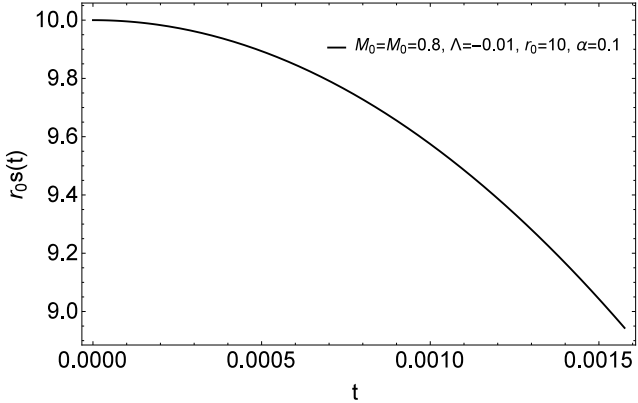}\caption{Evolution of radius $r_0 s(t)$} \label{fg12btz}
\end{figure}

\begin{figure}[tb]
\includegraphics[width=\columnwidth]{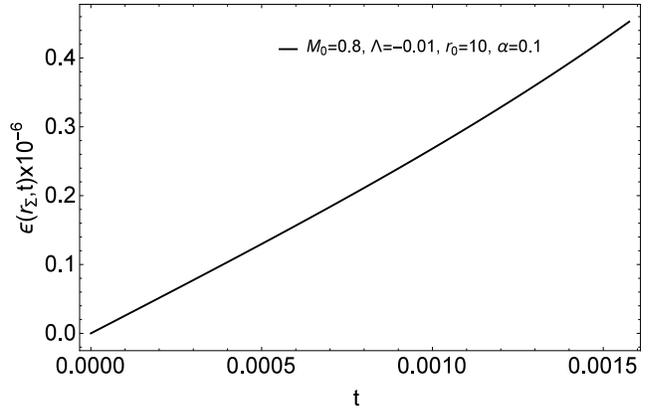}\caption{Evolution of radiation density $\epsilon(r_{\Sigma},t)$} \label{fg11btz}
\end{figure}

\begin{figure}[tb]
\includegraphics[width=\columnwidth]{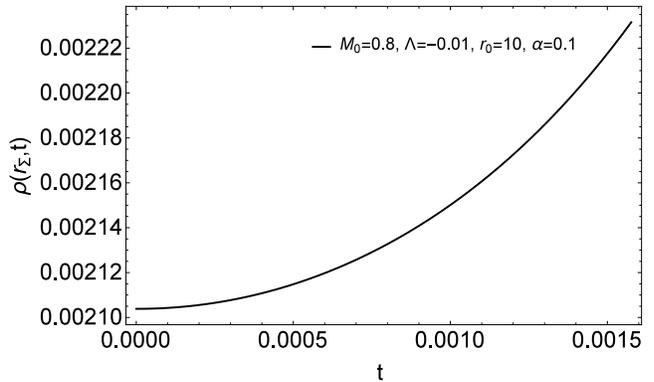}\caption{Evolution of surface density $\rho(r_{\Sigma},t)$} \label{fg13btz}
\end{figure}

\item {\bf Case II:}\\ 
Next, we choose $M_0 = 0.2$, $r_0 = 5$ and $\Lambda = -0.02$. For assumed value of $\alpha =0.3$, the constants are then calculated as $A = 0.0482$, $\beta = -0.0016$ and $\delta = 9.7497$. If we choose $\alpha =0.36$, the constants take the values $A = 0.0482$, $\beta = -0.002$ and $\delta = 11.8201$. We have compared evolution of physical quantities  for two different EOS parameter $\alpha$ in Fig.~(\ref{fg1btz})-(\ref{fg5btz}). The plots show the impact of the EOS parameter $\alpha$ on the collapse processes. We note that for a comparatively softer EOS, it takes more time to approach the horizon radius. This is also evident from the collapse rate 
$$\Theta = u_{;\beta}^{\beta} = \frac{3 \dot{s}}{s} e^{-\gamma_0}$$
for two different values of $\alpha$ as shown in Fig.~(\ref{fg6btz}).  

It should stressed here that the initial configuration, in our construction, is expected reach its horizon radius $r_h=\sqrt{-\frac{M_0}{\Lambda}} = 3.16228$ at times $t_h=$ $0.1412$ and $0.0201$ for assumed values of $\alpha =$ $0.3$ and $0.36$, respectively. However, it turns out that the total mass approaches zero  even before the shrinking boundary reaches its horizon radius in both the cases. Similar results, in the context of ($3+1$)-dimensional gravitational collapse, have been reported in some earlier works \citep{Banerjee,Chan3,Sharma2}. 

\end{itemize}

\begin{figure}[tb]
\includegraphics[width=\columnwidth]{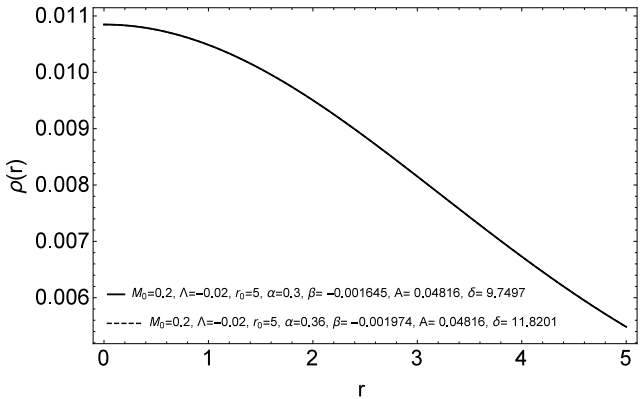}\caption{Energy density profile of the initial static configuration} \label{fg7btz}
\end{figure}

\begin{figure}[tb]
\includegraphics[width=\columnwidth]{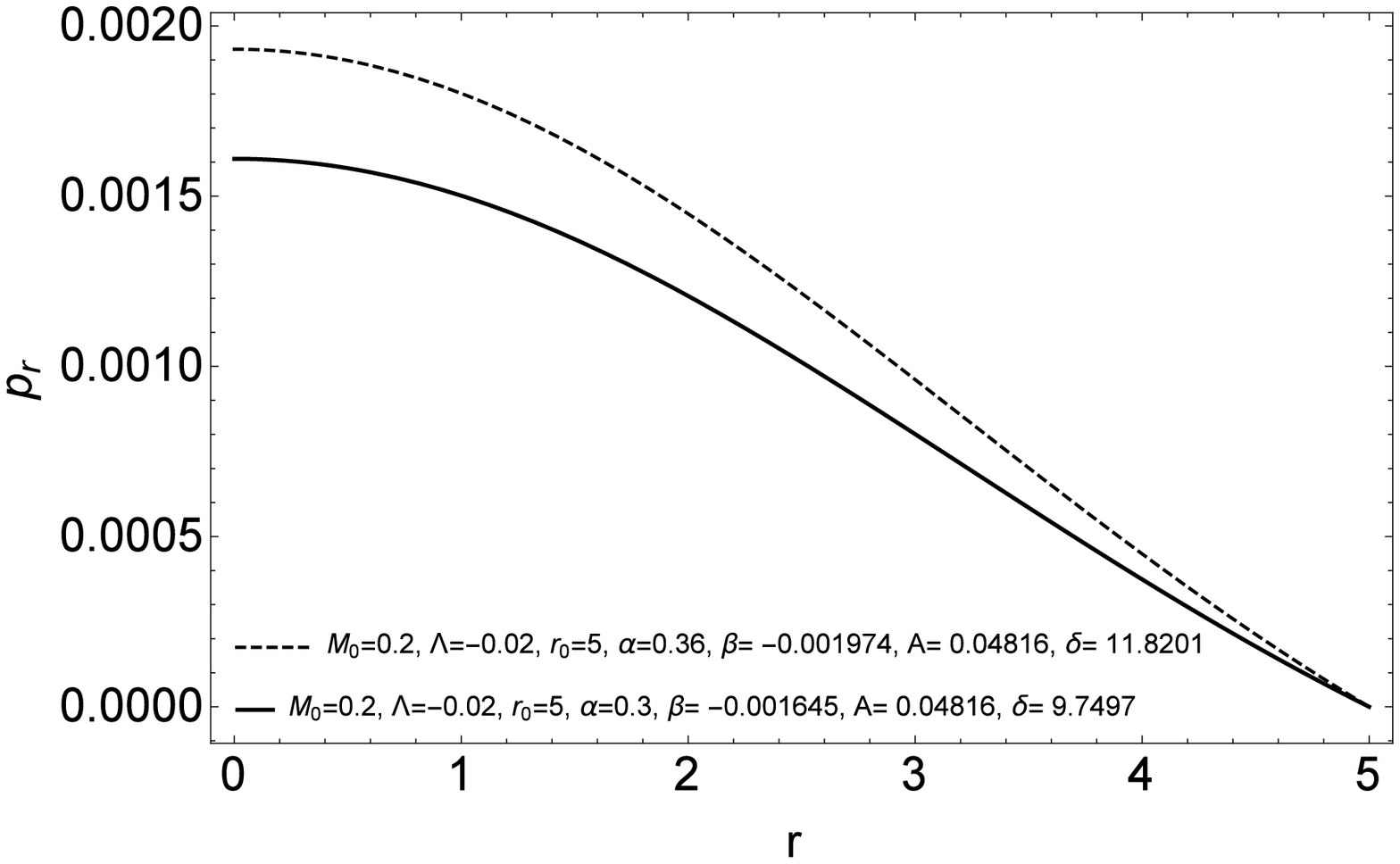}\caption{Radial pressure profile of the initial static configuration} \label{fg8btz}
\end{figure}

It's interesting to note that the collapse end state in this case is neither a black hole nor a naked singularity. In fact, the mass of the collapsing configuration tends to zero indicating that all the energy is radiated away even before the formation of the apparent horizon. We have checked that the initial static configuration is well behaved and the energy conditions are not violated. In Fig.~(\ref{fg7btz}) and (\ref{fg8btz}), we have shown that the respective energy-density and the radial pressure remain positive throughout the configuration which suggests fulfillment of (at least) the weak energy condition. 

\section{\label{sec7} Concluding remarks}
In this paper, we have analyzed gravitational collapse of a $3$-dimensional star emanating radiation whose exterior spacetime is described by the non-static generalization of the uncharged and spin-zero BTZ metric. In our model, the collapse begins from an initial configuration which is well-behaved and matter content of the initial configuration admits a linear equation of state. We have successfully established a link between the collapse rate and the EOS parameter.  By considering different initial configurations, we have also investigated the role of initial conditions on the collapse processes. Our results show that, for a given initial configuration and in the presence of a negative cosmological constant, the collapse eventually terminates with the formation of a BTZ black hole (Case I). However, we have also pointed out the possibility in which the mass of the collapsing star might disappear completely due to radiation even before the formation of the apparent horizon (Case II). Though, under what circumstances, all the energy gets evaporated rather than forming a black hole is not obvious, our studies clearly indicate the role of the initial conditions on the final fate of the collapse.  
 
To conclude, our approach brings out to attention the role of initial conditions on the evolution of a circularly symmetric collapsing matter in the presence of radiation. In the context of cosmic censorship conjecture \citep{Joshibook}, we note that while the initial configuration studied in Case I terminates into a black hole, it is not so in Case II. However, this does not indicate that the collapsing matter terminates into a naked singularity. In fact, even before the formation of the apparent horizon, the total mass of the particular collapsing configuration tends to zero as collapse progresses. It is to be stressed that the assumptions made in our approach to construct an artificial collapsing model are physically reasonable and mathematically coherent. Further, our results are in agreement with the observations made in more realistic ($3+1$)-dimensional collapsing models studied in Ref.~\citep{Banerjee,Chan3,Sharma2}.

\section*{Acknowledgements}
\noindent RS and FR gratefully acknowledge support from the Inter-University Centre for Astronomy and Astrophysics (IUCAA),
Pune, India, where a part of this work was carried out under its Visiting Research Associateship Programme.


\begin{thebibliography}{99}
\bibitem[\protect \citeauthoryear{Joshi}{1993}]{Joshibook} P. S. Joshi, {\it Global Aspects in Gravitation and Cosmology}, Clarendon Press, Oxford (1993).
\bibitem[\protect \citeauthoryear{Thorne}{1972}]{Thornebook} K. P. Thorne, {\it Magic Without Magic: John Archibald Wheeler}, edited by J. Klauder, Frimann, San Francisco (1972).
\bibitem[\protect \citeauthoryear{Virbhadra and Ellis}{2002}]{Vir1} K. S. Virbhadra and G. F. R. Ellis, {\it Phys. Rev. D} {\bf 65} (2002) 103004. 
\bibitem[\protect \citeauthoryear{Virbhadra and Keeton}{2008}]{Vir2} K. S. Virbharda and C. R. Keeton, {\it Phys. Rev. D} {\bf 77} (2008) 124014. 
\bibitem[\protect \citeauthoryear{Oppenheimer and Snyder}{1939}]{OppenS} J. R. Oppenheimer and H. Snyder, {\it Phys. Rev.} {\bf 56} (1939) 455.
\bibitem[\protect \citeauthoryear{Vaidya}{1953}]{Vaidya} P. C. Vaidya, {\it Proc. Indian Acad. Sci.} {\bf A33} (1951) 264 ; {\it Nature} {\bf 171} (1953) 260.
\bibitem[\protect \citeauthoryear{Santos}{1985}]{Santos} N. O. Santos, {\it Mon. Not. R. Astron. Soc.} {\bf 216} (1985), 403.
\bibitem[\protect \citeauthoryear{Joshi and Malafarina}{2011}]{Joshi} P. S. Joshi and D. Malafarina, {\it Int. J. Mod. Phys. D} {\bf20} (2011) 2641.
\bibitem[\protect \citeauthoryear{Ba\~{n}ados {\it et al}}{1992}]{BTZ} M. Ba$\tilde{n}$ados, C. Teitelboim and J. Zanelli, {\em Phys. Rev. Lett.} {\bf69} (1992) 1849.
\bibitem[\protect \citeauthoryear{Carlip}{1995, 1998}]{Carlip} S. Carlip, {\it Class. Quantum Grav.} {\bf12} (1995) 2853; {\it J. Korean Phys. Soc.} {\bf 28} (1995) S447; {\it Quantum Gravity in $2+1$ Dimensions}, Cambridge University Press, Cambridge, (1998).
\bibitem[\protect \citeauthoryear{Cruz and Zanelli}{1995}]{Cruz} N. Cruz and J. Zanelli, {\em Class. Quantum Grav.} {\bf12} (1995) 975.
\bibitem[\protect \citeauthoryear{Cruz {\it et al}}{2005}]{Cruz2} N. Cruz, M. Olivares and J. R. Villanueva, {\em Gen. Relativ. Grav.} {\bf37} (2005) 667.
\bibitem[\protect \citeauthoryear{Peixoto and Katanaev}{2007}]{Berredo} G. de Berredo-Peixoto and M. O. Katanaev, {\em Phys. Rev. D} {\bf75} (2007) 024004.
\bibitem[\protect \citeauthoryear{S\'{a}}{1999}]{Paulo} P. M. S$\acute{a}$, {\em Phys. Lett.} {\bf B467} (1999) 40.
\bibitem[\protect \citeauthoryear{Sharma {\it et al}}{2011}]{Sharma} R. Sharma, F. Rahaman and I. Karar, {\em Phys. Lett. B} {\bf704} (2011) 1.
\bibitem[\protect \citeauthoryear{Garc\'{i}ýa and Campuzano}{2003}]{Garc} A. A. Garc\'{i}ýa and C. Campuzano, {\em Phys. Rev. D} {\bf67} (2003) 064014.
\bibitem[\protect \citeauthoryear{Banerjee {\em et al}}{2013}]{Ayan} A. Banerjee, F. Rahaman, K. Jotania, R. Sharma and I. Karar, {\it  Gen. Relativ. Grav.} {\bf 45} (2013) 717.
\bibitem[\protect \citeauthoryear{Rahaman {\it et al}}{2013}]{Farook} F. Rahaman, P. K. F. Kuhfittig, B. C. Bhui, M. Rahaman, S. Ray and U. F. Mondal, {\it Phys. Rev. D} {\bf 87}, (2013) 084014.
\bibitem[\protect \citeauthoryear{Cataldo and Salgado}{1996}]{Cataldo} M. Cataldo and P. Salgado, {\it Phys. Rev. D} {\bf 54} (1996) 2971.
\bibitem[\protect \citeauthoryear{Garcia}{2004}]{Garcia} A. A. Garcia, {\it Phys. Rev. D} {\bf 69} (2004) 124024.
\bibitem[\protect \citeauthoryear{Banerjee {\em et al}}{2015}]{Ayan2} A. Banerjee, F. Rahaman, K. Jotania, R. Sharma and M. Rahaman, {\it  Astrophys. Space Sci.} {\bf 355} (2015) 353.
\bibitem[\protect \citeauthoryear{Rahaman {\it et al}}{2015}]{Farook2015} F. Rahaman, P. Bhar, R. Sharma and R. K. Tiwari, {\it Eur. Phys. J. C} {\bf 75} (2015) 107.
\bibitem[\protect \citeauthoryear{Mann and Ross}{1993}]{RossMann} R. B. Mann and S. Ross, {\it Phys. Rev. D} {\bf 47} (1993) 3319.
\bibitem[\protect \citeauthoryear{Mann and Oh}{2006, 2008}]{MannJohn} R. B. Mann and J. J. Oh, {\it Phys. Rev. D} {\bf 74} (2006) 124016; Erratum-ibid {\bf 77} (2008)  129902(E).
\bibitem[\protect \citeauthoryear{Gutti}{2005}]{Gutti} S. Gutti, {\it Class. Quantum Grav. } {\bf22} (2005) 3223.
\bibitem[\protect \citeauthoryear{Gutti and Singh}{2007}]{Gutti2} S. Gutti and T. P. Singh, {\it Phys. Rev. D} {\bf76} (2007) 064026.
\bibitem[\protect \citeauthoryear{Martins {\textit et al}}{2010}]{Martins} M. R. Martins, M. F. A. da Silva and A. D. Sheng, {\it Gen. Relativ. Grav.} {\bf 42} (2010) 281.
\bibitem[\protect \citeauthoryear{Miguelote}{2004}]{Miguelote} A. Y. Miguelote, N. A. Tomimura and A. Wang, {\it Gen. Relativ. Grav.} {\bf 36} (2004) 1883.
\bibitem[\protect \citeauthoryear{Chan {\textit et al}}{2006}]{Chan2} R. Chan, M. F. A. da Silva and J. F. V. da Rocha, {\it Int. J. Mod. Phys. D} {\bf15} (2006) 545. 
\bibitem[\protect \citeauthoryear{Choptuik}{1993}]{Choptuik1} M. W. Choptuik, {\it Phys. Rev. Lett.} {\bf 70} (1993) 912.
\bibitem[\protect \citeauthoryear{Ghosh}{2012}]{Ghosh} S. G. Ghosh, {\it Int. J. Mod. Phys. D} {\bf 21} (2012) 1250022.
\bibitem[\protect \citeauthoryear{Lemos}{1999}]{Lemos} Jos$\acute{e}$ P. S. Lemos, {\it Phys. Rev. D} {\bf59} (1999) 044020.
\bibitem[\protect \citeauthoryear{Govender and Thirukkanesh}{2009}]{Govender} M. Govender and S. Thirukkanesh, {\it Int. J. Theor. Phys.} {\bf48} (2009) 3558.
\bibitem[\protect \citeauthoryear{Dadhich and Ghosh}{2001}]{Dadhich} N. Dadhich and S. G. Ghosh, {\it Phys. Lett. B} {\bf518} (2001) 1.
\bibitem[\protect \citeauthoryear{Virbhadra}{1995}]{Virbhadra} K. S. Virbhadra, {\it Pramana-j. of phys.} {\bf44} (1995) 317.
\bibitem[\protect \citeauthoryear{Hussain}{1994}]{Hussain1} V. Hussain, {\it Phys. Rev. D} {\bf 50} (1994) R2361.
\bibitem[\protect \citeauthoryear{Hussain}{1995}]{Hussain2} V. Hussain, {\it Phys. Rev. D} {\bf52} (1995) 6860.
\bibitem[\protect \citeauthoryear{Vagenas}{2003}]{Vagenas} E. C. Vagenas, {\it Int. J. Mod. Phys. A} {\bf 18} (2003) 5949.
\bibitem[\protect \citeauthoryear{Israel}{1966}]{Israel} W. Israel, {\it Nuovo Cimento B} {\bf44} (1966) 1.
\bibitem[\protect \citeauthoryear{Banerjee {\textit et al}}{2002}]{Banerjee} A. Banerjee, S. Chatterjee and N. Dadhich, {\it Mod. Phys. Lett. A } {\bf17}, (2002) 2335.
\bibitem[\protect \citeauthoryear{Pinheiro and Chan}{2011}]{Chan3} G. Pinheiro and R. Chan, {\it Gen. Relativ. Grav.} {\bf43} (2011) 145. 
\bibitem[\protect \citeauthoryear{Sharma {\em et al}}{2015}]{Sharma2} R. Sharma, S. Das and R. Tikekar, {\it Gen. Relativ. Grav.} {\bf47} (2015), 25. 
\end{thebibliography}
\end{document}